\documentclass[12pt]{iopart}
\usepackage{amssymb}
\usepackage{epsfig}

\begin{document}

\title[Granular Freezing/Melting Transition]{Characterization of a Freezing/Melting Transition in a Vibrated and Sheared Granular Medium}
\author{Karen E. Daniels$^{1,2}$ \& Robert P. Behringer$^1$}
\address{$^1$Department of Physics and Center for Nonlinear and Complex Systems Duke University, Durham, NC, 27708, USA}
\address{$^2$ Department of Physics, North Carolina State University, Raleigh, NC, 27695, USA}
\eads{\mailto{karen\_daniels@ncsu.edu} and \mailto{bob@phy.duke.edu}}
\date{\today}

\begin{abstract}
We describe experiments on monodisperse spherical particles in an annular cell geometry, vibrated from below and sheared from above. This system shows a freezing/melting transition such that under sufficient vibration a crystallized state is observed, which can be melted by sufficient shear. We characterize the hysteretic transition between these two states, and observe features reminiscent of both a jamming transition and critical phenomena.
\end{abstract}

\submitto{Journal of Statistical Mechanics}

\maketitle

\section{Introduction}

Granular materials exhibit phases analogous to conventional solids,
liquids, and gases, in spite of being athermal and dissipative
\cite{Jaeger-1996-GSL}. Due to the dissipation, energy must be
supplied in order to sustain a dynamical state. Shearing and
vibration are two common means to inject energy into granular
systems. Shearing a granular material can compact and crystallize it
\cite{Tsai-2003-IGD}, but also melt it \cite{Daniels-2005-HCB};
tapping will compact it \cite{Nowak-1997-RIP}; in thin vibrated layers
there can be coexistence of crystallized and disordered states
\cite{Prevost-2004-NTP}; and highly vibrated granular systems become
gas-like. From a large phase space of variables we vary only two,
the shear rate and vibration amplitude, and study the interaction of
the two energy injection mechanisms.

Without vibration, sheared granular materials undergo a phase
transition from solid-like to fluid-like behavior: the particles must
become unjammed (which typically involves dilation) before they
can move. We seek to understand what effects vibrations have on such
transitions, and on the characteristics of the states on either side
of the transition. This is particularly interesting given that
granular systems are athermal, and one might naively expect that
vibrations would play a temperature-like role.

\begin{figure}[b]
\centerline{\epsfig{file=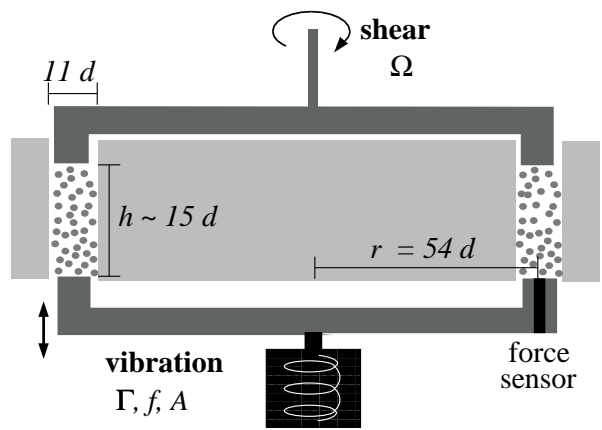, width=0.5\linewidth}}
\caption{Schematic cross-section of experiment (not to scale).}
\label{f_exp}
\end{figure}

We perform experiments in a classic geometry, annular shear flow
\cite{Savage-1984-SDD,Miller-1996-SFC,Losert-2000-PDS,Mueth-2000-SGM},
with monodisperse particles, shown schematically in Figure~\ref{f_exp}.
Shear and vibration provide competing
effects, with the system evolving to a crystallized state when the
kinetic energy provided by the vibration is greater than that provided
by the shear. The transition is hysteretic, and fluctuations in the
packing fraction and the breadth of the force distribution both become
large as the crystallized state is approached, in similarity to phase
transitions in other systems.

The physical parameters that characterize the system include the
amplitude $A$ and frequency $f$ of vibration, the height $H$
and mean radius $r$ of the annular container, the diameter $d$ and
the density $\rho$ of the particles, the rotation rate $\Omega$ of
the upper shearing surface, and the mean pressure $P$ on the layer
(here characterized at the base of the layer). From
these physical parameters, it is possible to define other dimensioned
parameters, such as the shear rate, $\dot{\gamma} = \Omega r/h$, as
well as a number of dimemsionless parameters which we list in
Table~\ref{t_dimnums}.

\begin{table}
\begin{center}
\begin{tabular}{c c l }\hline\smallskip
$\Gamma$ & $\displaystyle\frac{A \omega^2}{g}$ & vibrational acceleration / gravitational acceleration \\\smallskip
$H$ & $\displaystyle \frac{h}{d}$ & cell height / particle size \\\smallskip
$I$ & $\displaystyle\frac{\dot \gamma d}{c}$ & particle scale velocity / acoustic velocity \\\smallskip
$J$ & $\displaystyle\frac{\Omega r}{A \omega}$ & apparatus scale velocity / vibration velocity \\\smallskip
$K$ & $\displaystyle\frac{A \omega}{c}$ & vibration velocity / acoustic velocity \\\smallskip
$L$ & $\displaystyle\frac{A}{d}$ & vibration length scale / particle length scale \\\smallskip
$M$ & $\displaystyle\frac{\dot \gamma}{\omega}$ & shear time scale / vibration time scale \\\smallskip
$N$ & $\displaystyle\frac{P}{\rho g d }$ & applied pressure / hydrostatic pressure \\\smallskip
$R$ & $\displaystyle\frac{r}{d}$ & cell radius / particle diameter \\\smallskip
$\tilde \Omega$ & $\displaystyle\frac{\Omega r}{\sqrt{gd}}$ & shear velocity / particle velocity \\\hline
\end{tabular}
\end{center}
\caption{Dimensionless ratios, with $\omega \equiv 2 \pi f$, $\dot \gamma \equiv \Omega r/ h$, and $c \equiv \sqrt{P/\rho}$ \label{t_dimnums}}
\end{table}

$\Gamma$, $I$ and $J$ are three key parameters from the list in
Table~\ref{t_dimnums}. Of the ten listed, there are only seven
independent parameters: for example, $\Gamma = K^2 N / L$ and the
four velocity ratios $(H,I,J,K)$ only represent three parameters. In
the experiments described here, we have fixed $f$, $P$, and $N$
and therefore only explored a small region of the available phase
space.

\section{Experiment}

The experimental apparatus consists of an annular region containing
nearly monodisperse polypropylene spheres of diameter $d = 2.29$ to
2.39 mm and density $\rho = 0.90$ g/cm$^3$,
 as shown in Figure~\ref{f_exp}, with the pressure
$P$ and volume $V$ (height $h$) set from below by a spring within an
electromagnetic shaker. The particles are sheared from above and
vibrated from below, while the sidewalls are stationary. A more
detailed description of the apparatus is given in
\cite{Daniels-2005-HCB}. To characterize the states, we obtain
high-speed video images of particles at the outer Plexiglas wall,
laser position measurements of the bottom plate (cell volume), and
force time series from a capacitive sensor flush with the bottom
plate. For the experiments described in this paper, we fix the
frequency of vibration ($f=60$ Hz) and number of particles ($N \approx
71200$), and vary the amplitude of vibration $A$ and shear rate
$\Omega$. We vary the nondimensionalized peak acceleration
$\Gamma \equiv A (2 \pi f)^2/g$ from $0$ to $6$, and nondimensional
shear rate $\tilde\Omega \equiv \Omega r / \sqrt{gd}$
from $0.058$ to $9.3$.

\section{Description of States} %=================================
\begin{figure}[t]
\centerline{\epsfig{file=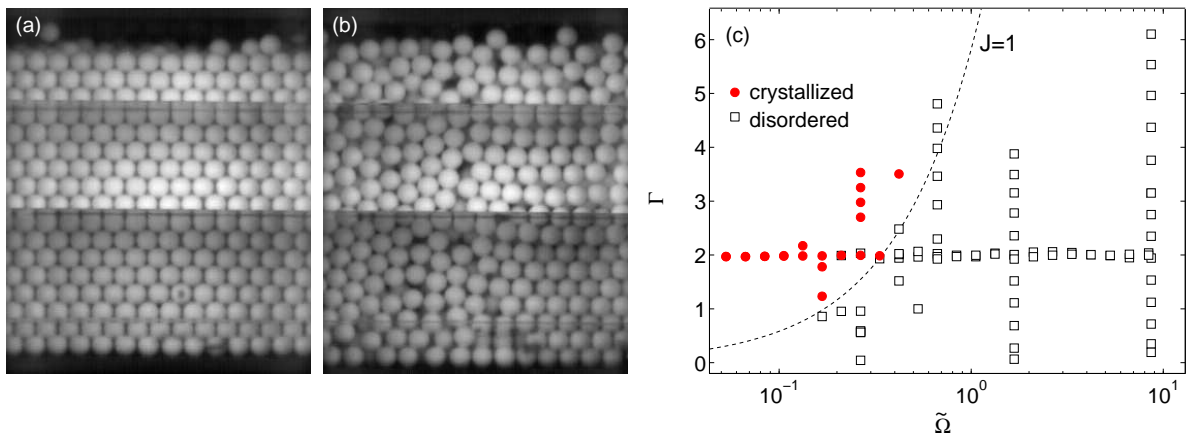, width=\linewidth}}
\caption{Sample images, viewed from outer wall. (a) Crystallized
state: $\Gamma=2.0$ and $\tilde \Omega = 0.078$. Movie at http://nile.physics.ncsu.edu/pub/cryst.mpg
(b) Disordered state: $\Gamma=2.0$ and $\tilde \Omega =
0.47$. Linear (L) and hexagonal (H) clusters marked by black
boxes. Movie at http://nile.physics.ncsu.edu/pub/disord.mpg
(c) Phase diagram for crystallized and disordered states as a
function of $\tilde\Omega$ and $\Gamma$. Dashed line is $J=1$.
Adapted from \protect\cite{Daniels-2005-HCB}.}
\label{f_phases}
\end{figure}

In the regime $0<\tilde\Omega<10$ and $0<\Gamma<6$ we observe two
distinct granular states of matter: crystallized and
disordered. Sample images and movies of these two states are shown in
Figure~\ref{f_phases}ab, as viewed from the outer wall. For $\tilde
\Omega \lesssim 1$ and $\Gamma < 4$, we observe that the phase
boundary between the two states roughly corresponds to a curve where
the characteristic velocities of the two motions are equal. This corresponds
to the dimensionless number $J$ of Table~\ref{t_dimnums}:
\begin{equation}
J \equiv \frac{\Omega r}{2 \pi f A}
\label{e_Jdef}
\end{equation}
and the curve $J=1$ is shown by the dashed line in
Figure~\ref{f_phases}c. Below, we characterize these two states, with
further details to be found in \cite{Daniels-2005-HCB}.

\begin{figure}
\centerline{\epsfig{file=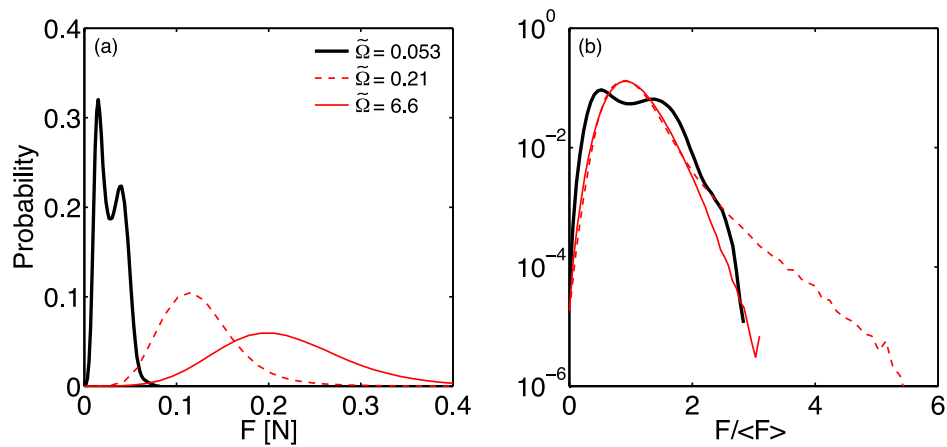, width=0.8\linewidth}}
\caption{Force probability distribution functions for three values of
$\tilde \Omega$ at $\Gamma = 2.0$. (a) on linear scales,
dimensioned and (b) on log-linear scales, normalized by mean force.
\label{f_fpdfs}}
\end{figure}

{\itshape Crystallized State:} In the solid-like state (see
Figure~\ref{f_phases}a), the balls crystallize into a hexagonally
close-packed configuration ({\em i.e.} a 3D crystalline structure)
here visible only at the outer wall although the order persists across
the layer. The contact between the upper layer of the granular
material and the shearing wheel is intermittent, with stick-slip
motion of the top $\sim2$ layers in the manner of
\cite{Nasuno-1997-FGL}. The distribution of forces measured at the
bottom of the layer is bimodal (see Figure~\ref{f_fpdfs}); this
indicates that the material is responding as a solid body moving up
and down with the sinusoidal vibrations of the bottom plate.

{\itshape Disordered State:} In the disordered state, some order
remains in the form of hexagonally-packed clusters and linear chains
of particles at the outer wall, as marked in
Figure~\ref{f_phases}b. For states with $\Omega$ well above the
transition, linear chains dominate over hexagonal clusters, with both
existing intermittently throughout the disordered regime. These chains
may correspond to the planar ordering reported recently by Tsai et
al. \cite{Tsai-2003-IGD,Tsai-2004-SSD}. The velocity profile extends
deeper into the layer (in the vertical direction) than in the
crystallized state. Force distributions measured at the bottom plate
show the exponential-like tails characteristic of many granular
experiments in disordered, unvibrated granular materials (see
Figure~\ref{f_fpdfs}). They also fall to zero at low force, as seen
in earlier experiments by Miller et al. \cite{Miller-1996-SFC}.

For a geometrically similar system, but unvibrated and exposed to a
compressional force, shear {\em ordered} the system into horizontal
planes of hexagonal packing, each slipping past the others
\cite{Tsai-2003-IGD,Tsai-2004-SSD}. Such a state is different from the
3D crystallized state observed here, in which the layers in the bulk
are stationary with respect to each other. An interesting question is
how shearing creates order or disorder depending on the
presence or absence of vibration. A useful way to distinguish the
ordered and disordered regimes is via the ratio
\begin{equation}
I \equiv \frac{\dot \gamma d}{\sqrt{P/\rho}}
\label{e_pouliquen}
\end{equation}
involving the shear time scale $\dot \gamma \equiv \Omega r / h$ to
the acoustic time scale, $\sqrt{d^2 \rho /P}$, calculated from the
pressure $P$ and density $\rho$ \cite{Midi-2004-DGF}. The experiments
described in this paper have pressures of around $P=20$ Pa and shear
rates of $\dot \gamma = 0.3$ to 40 Hz, leading to $I = 5 \times
10^{-3}$ to 0.75. In \cite{Tsai-2003-IGD}, the shear rates are slower,
$\dot \gamma = 0.05$ to 0.5 Hz, and pressures are higher, $P = 2000$
Pa, so that the system is clearly in the quasistatic regime with $I= 3
\times 10^{-5}$ to $3 \times 10^{-4}$.

While it is perhaps surprising that we find as simple a result
as a phase transition at $J\approx 1$, the presence of these other
important control parameters give hints into the breakdown of
crystallization for large $\Gamma$. Figure~\ref{f_phases}c shows that
for $\tilde \Omega \approx 0.7$, crystallization was not observed
above $\Gamma = 4$, possibly indicating the re-emergence of disorder
due to granular-gas like behavior. Further experiments varying
$(L,M,N)$ will be necessary to discover which of these determine
the high-$\Gamma$ boundary of the crystallized phase. In addition, the
parameter $H$ controls finite size effects.

\section{Shear Localization}

\begin{figure}
\centerline{\epsfig{file=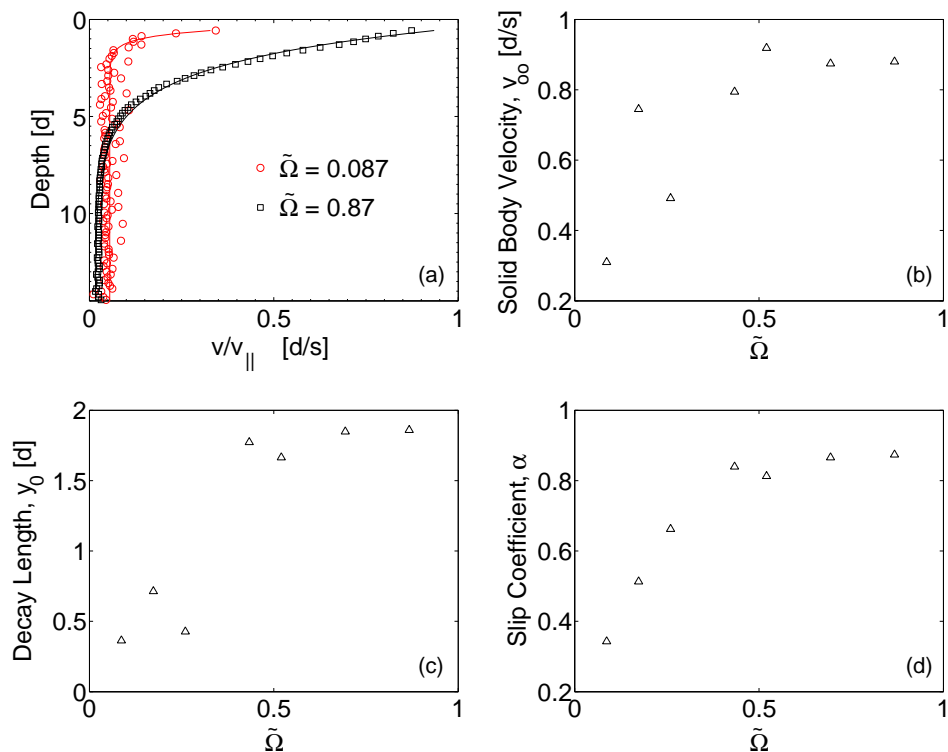, width=0.8\linewidth}}
\caption{Characterization of velocity profiles at $\Gamma = 2.0$ and various $\tilde \Omega$. (a) Azimuthal velocity measured at outer wall as a function of depth measured. Lines are fits to Equation (\ref{e_velfit}). (b) Solid body rotation $v_\infty$ as a function of $\tilde \Omega$, (c) decay length $y_0$ as a function of $\tilde \Omega$ and (d) slip coefficient $\alpha$ as a function of $\tilde \Omega$. \label{f_rotvelprofile}}
\end{figure}

Granular materials commonly exhibit shear banding, with an exponentially decaying velocity profile away from the shearing surface. As shown in Figure~\ref{f_rotvelprofile}a, this shear band behavior is seen in both the crystallized $(\tilde \Omega = 0.087)$ and disordered states $(\tilde \Omega = 0.87)$ for $\Gamma = 2.0$. To obtain the velocity profiles, we tracked individual particles visible at the outer wall using a high-speed video camera and determimed trajectories for each. We then binned the resulting velocity components by depth to construct velocity profiles.

We characterize the azimuthal velocity, $v(y)$, by fitting the profile to the form
\begin{equation}
v(y) = v_\infty + \alpha\, v_\parallel\, e^{-y/y_0}
\label{e_velfit}
\end{equation}
where $v_\infty$ is the solid-body motion at the bottom plate, $v_\parallel$ is the known azimuthal velocity of the shear wheel at the outer wall, $\alpha$ is the efficiency with which that velocity is transmitted to the top layer of granular material, and $y_0$ is the decay length of the velocity with depth.

In the crystallized state, the shear is localized almost entirely to
the first layer of particles (small $y_0$), while in the disordered
state the shear band extends several particles into the layer. The
slip at the upper plate is lowest in the disordered state, where the
uppermost particles are in constant contact with the shearing wheel.
Note that the system is more dilated in the disordered state, and has
a larger pressure \cite{Daniels-2005-HCB}. While the disordered state
has greater slip ($v_\infty$) at the bottom plate, the scaled slip
values ($v_\infty/v_\parallel$) are in fact lower than in the
crystallized state, visible in Figure~\ref{f_rotvelprofile}a. For
disordered states with $\tilde \Omega \gtrsim 0.4$, the shear bands
appear to have reached a steady state since they are all parameterized
by the same values.

\section{Transition} %==============================================
\begin{figure}
\centering
\centerline{\epsfig{file=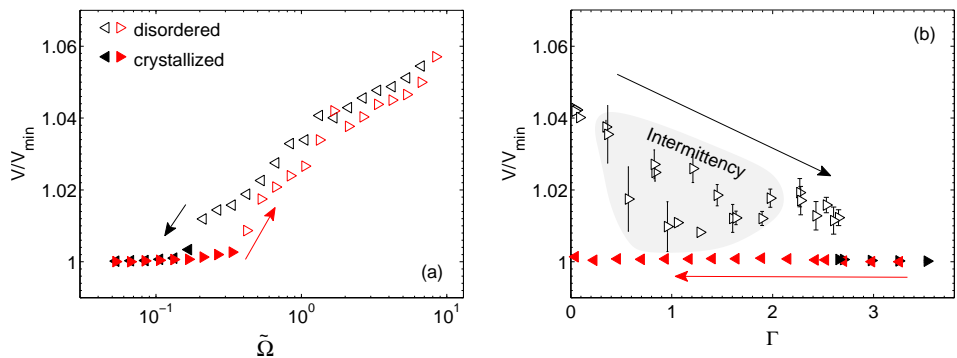, width=0.8\linewidth}}
\caption{Volume $V$ of cell, scaled by minimum observed volume
$V_{min}$ as a function of (a) $\tilde \Omega$ (at $\Gamma = 2.0$) and
(b) $\Gamma$ (at $\tilde \Omega = 0.27$). Triangles point in direction
of steps.}
\label{f_hysteresis}
\end{figure}

We examined the transition from the disordered to the crystallized
state by first preparing a disordered state at high $\tilde\Omega$,
then adjusting $\tilde \Omega$ to the value of interest. We then
performed two runs, one at constant $\tilde \Omega = 0.27$ (starting
from $\Gamma=0$) and the other at constant $\Gamma =2$ (starting from
$\tilde \Omega = 8.4$). The mean volume measured for each step of
these two runs is shown in Figure~\ref{f_hysteresis}.

For steps of decreasing $\Omega$ (Figure~\ref{f_hysteresis}a) the
system compacts logarithmically until reaching $\Omega_c$, after which
the system undergoes a first order phase transition to the
crystallized state. Below $\Omega_c$ only a small amount of additional
compaction occurs, to a state with a volume $V_{min}$, for which the
packing fraction is $\phi= 0.69$. When $\Omega$ is increased, the
transition back to the disordered state is hysteretic, occurring for
$\Omega_h \approx 2 \Omega_c$.

For steps of increasing $\Gamma$ (Figure~\ref{f_hysteresis}b) the
system also compacts. However, runs approaching the transition are
difficult to repeat quantitatively, since there is a great deal of
intermittency in the cell volume (see Figures~\ref{f_hysteresis}b,
\ref{f_velprofile}a and \cite{Daniels-2005-HCB} for details). For
$\Gamma > \Gamma_c$ the system is in the crystallized state. The
transition also appears to be first order, but in this case the
hysteresis is so extreme that the material was not observed to
re-expand when we decreased $\Gamma$.

We wish to understand why a crystallized state can disorder by
increasing $\Omega$, but not by decreasing $\Gamma$. In the case of
increasing $\Omega$, the stick-slip behavior in the top layers of the
crystallized state is affected by the speed of the upper shearing
wheel. As $\Omega$ increases, more horizontal momentum is transfered
to the upper layer of balls, which results in longer regions of
flowing particles. Eventually, the whole layer can be seen to mobilize
and the disordering begins to take place throughout the cell. In
contrast, for increases in $A$ (and hence $\Gamma$) no such increased
momentum transfer takes place, and the results are similar to the
irreversibility observed for compaction by tapping
\cite{Nowak-1997-RIP}. This transition shows some similarity
similarity to the ``freezing-by-heating'' transition seen in
\cite{Helbing-2000-FHD}, in which individual particles with tunable
noise are seen to crystallize as their noise level is increased. Such
a system also shows hysteresis in returning to the disordered,
mobilized state.

\begin{figure}
\centering
\centerline{\epsfig{file=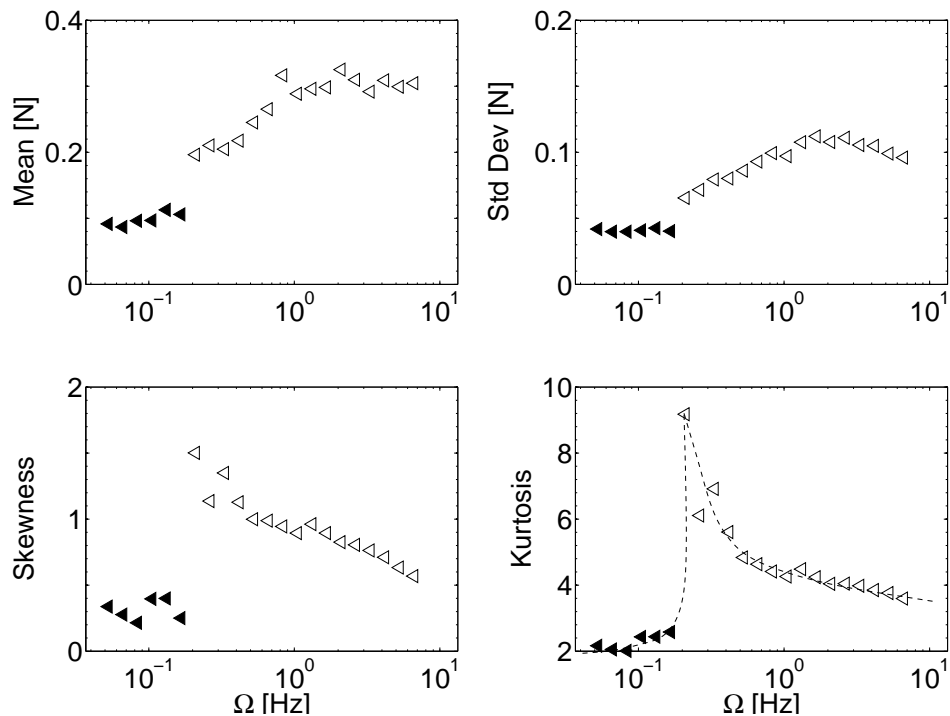, width=0.8\linewidth}}
\caption{Characteristics of force probability distributions as a function of $\tilde
\Omega$ at $\Gamma = 2.0$. Triangles point in direction of steps in
$\tilde \Omega$; solid points are crystallized phase; dashed line is
guide to the eye. \label{f_rotforce}}
\end{figure}

\begin{figure}
\centering
\centerline{\epsfig{file=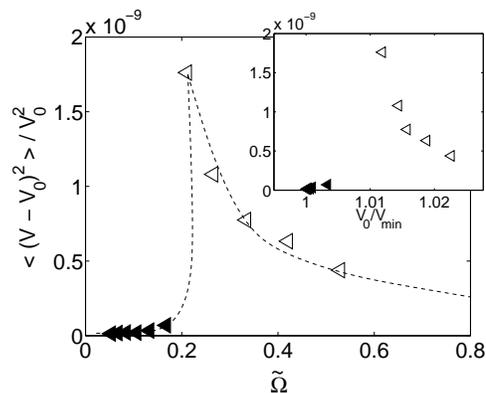, width=0.4\linewidth}}
\caption{Volume fluctuations as a function of $\tilde \Omega$, scaled
by average volume $V_0$. Inset: Volume fluctuations as a function of
$V_0/V_{min}$ at each $\tilde \Omega$. Triangles point in direction of
steps in $\tilde \Omega$; solid points are crystallized phase; dashed
line is guide to the eye. \label{f_rotfluct}}
\end{figure}

For the run at $\Gamma=2.0$, seen in Figure~\ref{f_hysteresis}a with
steps downward in $\Omega$, we observe signatures of the phase
transition from disorder to crystallization via both the volume
fluctuations and the force distribution, as shown in
Figures~\ref{f_rotforce} and \ref{f_rotfluct}. As $\Omega \rightarrow
\Omega_c$ from above, both the volume fluctuations (measured from the
variance of $V(t)$) and the breadth of the force distribution
(measured by the kurtosis, or fourth scaled moment, of $F(t)$ on the
force sensor) become large. In addition, the first order nature of the
transition is visible in other characteristics of the force
distribution, such as the mean, standard deviation, and skewness (see
also Figure~\ref{f_fpdfs}).

In granular systems, Edwards and coworkers \cite{Edwards-1989-TP} have
introduced a temperature-like measure, the compactivity, defined as
$X=(\partial V/\partial S)_N$ by analogy with thermodynamics. The
central idea is that lower packing fractions correspond to a greater
freedom for particle rearrangement, and hence a higher
compactivity. In the statistical mechanics of ordinary second order
phase transitions, susceptibilities such as $(\partial^2 A/\partial
T^2)_V$ (for free energy $A$) are singular at the critical point. For
example, the specific heat at constant volume is $C_V = (\partial E /
\partial T)_V = -T (\partial^2 A/\partial T^2)_V$. When described in
terms of fluctuation-dissipation relations, $k_B T^2 C_V = \langle
(E-E_0)^2 \rangle$, where $E$ is the energy of the system and $E_0$
its mean value. One expects energy fluctuations, and hence
$C_V$, to be singular at the critical temperature $T_c$. By
contrast, at a first order transition, discontinuities occur in
densities, but one does not expect divergent fluctuations. Since $V$
has taken the place of $E$ in the Edwards formalism, the hallmark of a
critical transition is increased fluctuations in the volume of the
system as we approach $X_c$, the critical compactivity. In our
experiments, volume (and hence $X$) is set by $\Omega$, and the inset
to Figure~\ref{f_rotfluct} shows apparently singular behavior for
the volume fluctuations as a function of the volume. It is
interesting that in these experiments, we see a discontinuity in the
density, but also an indication of a singularity in the volume
fluctuations. The magnitude of the fluctuations observed in the
disordered state is similar to those in observed in
\cite{Schroter-2005-SSV}, where the standard deviation of the packing
fraction is approximately $10^{-4}$.

\section{Intermittency}

\begin{figure}
\centerline{\epsfig{file=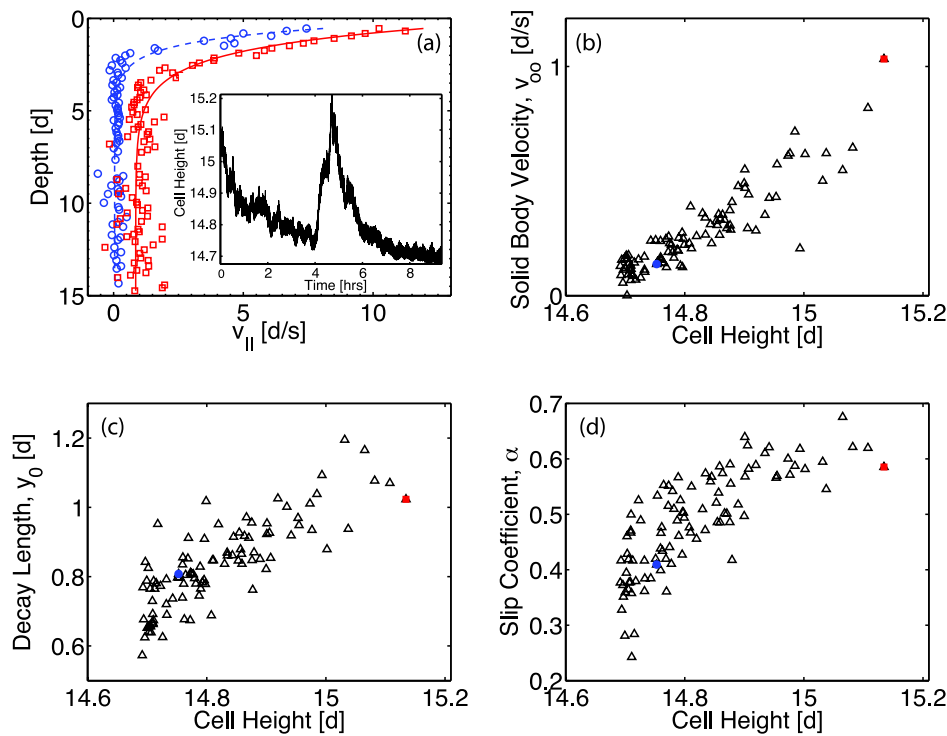, width=0.8\linewidth}}
\caption{Characterization of velocity profiles in intermittent regime
at $\tilde\Omega = 0.27$ and $\Gamma = 1.0$. (a) Azimuthal velocity
measured at outer wall as a function of depth measured for compact
($\bigcirc$, $t=3.8$ hrs) and dilated ($\square$, $t=4.7$ hrs)
states. Lines are fits to Equation (\ref{e_velfit}). (b) Solid body
rotation $v_\infty$ as a function of cell height, (c) decay length
$y_0$ as a function of cell height and (d) slip coefficient $\alpha$
as a function of cell height. Solid symbols correspond to data from
(a). \label{f_velprofile}}
\end{figure}

The apparently singular volume fluctuations near $J=1$ come from the
fact that the system exhibits intermittency in its state. The system
is in fact spatially inhomogeneous, with instances of small $V$ being
crystallized in the majority of the cell and instances of large $V$
being majority disordered. By examining the properties of the system
in this intermittent regime, we are able to compare a broad range of
states for nearly the same parameter values. The only varying
parameters are the volume and pressure of the system, which are
related to each other by a proportionality constant due to the spring
constant of the shaker \cite{Daniels-2005-HCB}.

The inset to Figure~\ref{f_velprofile}a shows a time series of the
cell height for such a run, with $\Gamma = 1.0$ and $\tilde \Omega =
0.27$. The system started from a dilated state and progressed to a
majority crystallized state before re-dilating and re-compacting to an
even more crystallized state over the course of approximately 10
hours. We again obtain velocity profiles at the outer wall, in this
case while simultaneously monitoring the position of the bottom plate.

Figure~\ref{f_velprofile}a shows two velocity profiles from compact
and dilated states. Because the system is spatially inhomogeneous, the
particles in view of the camera may in fact be either disordered or
crystallized at any given time during these measurements, regardless
of the height of the cell. Importantly, the fit parameters in
Figure~\ref{f_velprofile}b-d show the same trends as those in
Figure~\ref{f_rotvelprofile} when rotation rate $\tilde \Omega$ is
taken as a proxy for cell height. In both cases, we observe a
continuum of states as the system moves between crystallization
(compaction) and disorder (dilation). Again, it is interesting to
note that the volume and pressure fluctuations are associated with the
formation and melting of ordered clusters over time. Such behavior is
characteristic of near-critical behavior. By contrast, at a
thermodynamic first order transition, we would not expect to see
persistent dominant fluctuations.

\section{Discussion} %==============================================

The two characterizations of a transition in the system
we discuss above provide contrasting, but complimentary,
information about the nature of the crystallizing phase transition in
sheared and vibrated granular materials. The canonical hallmark of a
transition to a jammed/glassy state is the continuous growth of the
viscosity. Glass transitions do not in general contain
first-order-like signatures, such as discontinuities in the volume or
specific heat \cite{Torquato-2000-GTH}. For sheared colloids, there
are large stress fluctuations near a jamming transition
\cite{Lootens-2003-GSF}, and in simulations of Lennard-Jones
particles, force PDFs are observed to broaden
\cite{OHern-2001-FDN,Snoeijer-2004-FNE}. Similar behavior is observed
in this system as well, but with a density discontinuity. By
contrast, jammed/glassy states are all {\em disordered}, while the
granular system described in this paper makes a transition to a {\em
crystallized} state. In both the glassy and crystallized
cases, however, the final states are unable to rearrange.

We observe similarities to critical phenomena in the increased volume
fluctuations near the transition, a hallmark at odds with a glass
transition. These fluctuations are similar to the density fluctuations
observed at the liquid-gas critical point, which occur at diverging
length scales. Therefore, further investigations into the nature of
this transition should examine what length scales and order parameters
are present, including a determination of the sizes of clusters and
the spatial correlations between forces. Finally, we have introduced a
number of dimensionless control parameters whose effects remain to be
investigated in future studies.

\ack

This research was supported by the NASA Microgravity program under grant NNC04GB08G, and North Carolina State University.

%\bibliographystyle{unsrt}
%\bibliography{granular,ked}
\end{document}